\documentclass[9pt]{elife}

\usepackage{amsmath,amsfonts}
\usepackage[utf8]{inputenc}
\usepackage[T1]{fontenc}

\usepackage{todonotes}

\title{InsectSet459: an open dataset of insect sounds for bioacoustic machine learning}

\author[1]{Marius Fai\ss}
\author[2]{Burooj Ghani}
\author[2,3]{Dan Stowell}

\affil[1]{Max Planck Institute of Animal Behavior, Konstanz, Germany}
\affil[2]{Naturalis Biodiversity Centre, Leiden, The Netherlands}
\affil[3]{Tilburg University, Tilburg, The Netherlands}

\corr{mfaiss@ab.mpg.de}{MF}

\begin{document}


\maketitle

\begin{abstract}

Automatic recognition of insect sound could help us understand changing biodiversity trends around the world---but insect sounds are challenging to recognize even for deep learning. We present a new dataset comprised of 26399 audio files, from 459 species of Orthoptera and Cicadidae. It is the first large-scale dataset of insect sound that is easily applicable for developing novel deep-learning methods. Its recordings were made with a variety of audio recorders using varying sample rates to capture the extremely broad range of frequencies that insects produce. We benchmark performance with two state-of-the-art deep learning classifiers, demonstrating good performance but also significant room for improvement in acoustic insect classification. This dataset can serve as a realistic test case for implementing insect monitoring workflows, and as a challenging basis for the development of audio representation methods that can handle highly variable frequencies and/or sample rates.
\end{abstract}

\section{Introduction}

Insects are crucial members of many ecosystems, and so the recent reports of insect population declines are a vital global topic \citep{Wagner:2021}. Yet there is still much uncertainty about insect biodiversity and population changes, due to a poverty of monitoring information from a very limited number of species and geographic regions  \citep{Montgomery:2020}.
Automatic monitoring has a key role to play, including automatic recognition from sounds (and from images) \citep{vanKlink:2022,vanKlink:2024,Kohlberg:2024,Riede:2024}.

There is a vast number of soniferous insect species including large clades whose sounds carry species-specific information, with two of the biggest and loudest being Orthoptera and Cicadidae
\citep{Riede:2024}. To monitor population changes, distributions and rare species occurrence, automatic acoustic detection and classification methods can be applied in a wide range of environments while being minimally invasive \citep{Montgomery:2020} and automatable to a high degree \citep{Riede:2018}.
The scale of this task seems daunting, due to the sheer number of insect species, but on the other hand, modern deep learning benefits from large-scale data, and certain tasks become tractable only when a certain scale is attained.
For example, bird vocalization classifiers---trained on hundreds of species using deep learning---have now established themselves in the toolbox of bird monitoring techniques \citep{PerezGranados:2023,Sethi:2024}.

For soniferous insects, the data available for deep learning is not yet large-scale---neither in terms of audio duration, nor species coverage. In their systematic review
\cite{Kohlberg:2024} identified machine learning methods totaling 302 species across all studies.
In earlier work we introduced curated open datasets of insect sounds containing up to 66 species\footnote{\url{https://doi.org/10.5281/zenodo.7828438}} of Orthopterans and Cicadas \citep{Faiss:2023}. However, much larger collections are needed, in order for deep learning to become usefully deployable for acoustic insect monitoring.
In the same work we evaluated a novel feature representation, based on learnable filter banks instead of spectrograms. Various other machine learning strategies may be useful for insect sound recognition, but it is not yet clear whether the best insect recognition will be based on waveforms, spectrograms, adaptive filters, or other new types of representation that suit the acoustic phenomena. There is a need for more comprehensive data, both for development of new methods and for validation of the technical readiness level of classifiers.

In 2024, the public animal sound database xeno-canto\footnote{\url{https://xeno-canto.org}} witnessed a dramatic increase in insect sound recordings. This is due to the publication of several large collections of field and laboratory recordings from insect sound experts, as well as increased adoption of citizen scientists uploading their insect sound observations to the website.
At around the same time, the iNaturalist\footnote{\url{https://www.inaturalist.org}} project has also witnessed large growth in the number of sound recordings collected, including sound from insects \citep{Chasmai:2024}. These two projects have in common the collection of species-labeled sound recordings, an open data philosophy (under varying licenses including multiple Creative Commons licenses), and the use of citizen science (community science) to bring together sounds from around the world. Their focus on wildlife, unlike general-purpose audio collections, encourages the community to work towards correct species labeling and representative audio clips.

We used this opportunity to compile the first large-scale dataset of insect sounds that is easy to use for training deep learning methods to detect and classify insect sounds in the wild.
We note a recent parallel initiative to curate a general-purpose audio dataset from iNaturalist \citep{Chasmai:2024}.
Unlike that initiative, we focused on preparing the data for the specific goal of enabling insect sounds to be better recognized through machine learning, since this is a current important gap.

In this paper we introduce InsectSet459, an acoustic dataset of 459 insect species, curated from xeno-canto and iNaturalist open data and preprocessed to be of maximal usefulness to develop machine learning for insect sound.
We describe the dataset characteristics and the curation process. We also present results of applying deep learning algorithms for automatic species classification of insect sounds, demonstrating that the dataset enables high-quality acoustic recognition.
We intend that others can use the dataset to create even stronger classifiers than those we demonstrate, so that intelligent acoustic monitoring can contribute to the urgent task of insect population monitoring.

\section{Materials and methods}

\subsection{Curation of InsectSet459}

Our data curation was an expanded and revised version of the method of \cite{Faiss:2023}.
Recordings were downloaded and pooled together in October 2024 from the following three sources:
\begin{itemize}
    \item xeno-canto: Orthoptera
    \item iNaturalist: Orthoptera \& Cicadidae
    \item BioAcoustica \citep{Baker:2015}: Cicadidae
\end{itemize}
Initially, all files belonging to observations at the species-level were downloaded from xeno-canto and Bioacoustica. From iNaturalist, only research-grade observations were downloaded. For observations with multiple audio files attached, only one file was downloaded.
We exclude sound files with ``no derivatives'' license limitations to ensure re-usability of the dataset and to be able to publish shortened versions of very long audio files.
All recordings are licensed under creative commons licenses CC-BY-4.0 or CC0, which therefore means that users are permitted extensive rights to reuse, share and publish the resulting data.

Deduplication is an important curation step for machine learning, especially when multiple data sources are used.
If users uploaded to both iNaturalist and xeno-canto, only the files from one of the platforms were used. Further, we performed deduplication based on MD5 checksums to reduce the risk of duplicate files, especially considering that some files are in practice uploaded both to xeno-canto and to iNaturalist. We also found that some users had uploaded the same files multiple times to the same platform, sometimes listed as different species. Duplicate recordings uploaded under the same species name were removed, recordings that were uploaded under multiple species names were completely excluded from the data pool.
Another common source of redundancy is serial uploads from one location and time period split into separate observations (especially common on xeno-canto), which could include the same individual animals vocalizing. This problem was addressed by pooling all recordings by username, species, geographic location, date and time, and then selecting only one recording from within a one-hour period.

After these filtering steps, all files from species with at least 10 remaining sound examples were selected for the final dataset.
In our previous work, for InsectSet32\footnote{\url{https://doi.org/10.5281/zenodo.7072195}} we used sources with fewer recordings available, but divided those recordings into multiple sections. This is useful for creating enough training data for deep learning, but it does not yield a large diversity in the recording conditions, and thus it carries some risk of overfitting in training (a kind of ``false replication'' in experimental design). For the present work we restricted to 10+ separate recordings per species, to encourage that all of the classes that a classifier should learn are sufficiently representative and diverse.

All stereo files were converted to mono. File formats were standardized to WAV and MP3, both of which are present in much of the source material, and also very widely compatible with audio tools. Less-common formats such as M4A were converted to high-quality MP3 to ensure wide compatibility.

We did not reduce the sample rate down to a common value. This is a strategic difference from common practice in machine learning dataset work, and is motivated by the dramatic variation of bandwidths known in insect sound which can reach far outside of the human hearing range and can therefore not be represented at the most commonly used sample rates. We provide audio at the full sample rate given in the original source, which avoids information loss and allows for high-bandwidth and ultrasonic deep learning methods to be developed in the future.

Sound files come in dramatically varying durations, from less than one second to many minutes. In order to maximize the acoustic diversity for a given size of dataset, we chose to trim each audio file down to a maximum duration (2 minutes). In a long audio file, we skip the first 2 minutes (since it may contain speech or handling noise) and take the next 2 minute segment.
We did not manually trim the audio down to song bouts (echemes), because that is not feasible at large scale; instead we automatically trimmed to 2-minute segments. 
It risks making some imperfect edits, which could impede the training of algorithms. On the other hand it also corresponds more closely to automated passive acoustic monitoring, in which arbitrary sound segments are usually submitted for analysis. For those who wish to use as much audio data as possible, and thus more than 2 minutes per file, the original sources are available as links in the annotation file.

Species nomenclature was unified to the Orthoptera Species File taxonomy (COL24.4 2024-04-26 [294826]) using checklistbank \citep{Doring:2022}. 
Xeno-canto's metadata structure offers its users the possibility to  label species that are audible in the background of a recording alongside the main species. However, since most users do not enter this information, we decided not to exclude these recordings from InsectSet459. In practice, field recordings of insects commonly contain sounds and chorusing from other species in the background and they are usually not labeled, especially on iNaturalist. Therefore, excluding recordings from xeno-canto with background species labeled likely would not improve the overall quality of the dataset.  We instead add this information to the metadata file in a separate column, which could potentially be useful for users of the dataset in some cases.

For machine-learning purposes, the dataset was split into the training, validation and test sets while ensuring a roughly equal distribution of audio files and audio material for every species in all three subsets. This resulted in a 60/20/20 split (train/validation/test) by file number and file length (\TABLE{filedurations}).

This new dataset greatly increases the number of species compared against prior work, giving 459 unique species from the groups Orthoptera and Cicadidae, while also strongly increasing the geographic coverage of recording locations (\FIG{datasets_subsets_map}). The total duration of the dataset and number of sound examples is heavily expanded to a total of 26399 files containing 9.5 days of audio material (\TABLE{filedurations}) with sample rates ranging from 8 to 500 kHz (\TABLE{fileformats}).
The dataset is published on Zenodo\footnote{\url{https://doi.org/10.5281/zenodo.14056457}}, and also on AcademicTorrents\footnote{\url{https://academictorrents.com/details/a8278b49a33cc05a23b14aebd2da75d693012678}}.

\subsection{Classifier development: InsectEffNet and PaSST}

To provide a baseline of the recognition performance for insect sounds based on this dataset, following the current state of the art in acoustic machine learning, we trained two deep learning classifiers.
Although our dataset is designed to encourage innovative approaches including the use of diverse sampling rates, for the present work we restricted our attention to the performance of algorithms which use the standard spectrogram approach of audio classification \citep{Stowell:2022review} i.e.\ converting the waveform data to a fixed non-ultrasonic sampling rate, transforming them into Mel spectrograms. The resulting classification problem can be addressed using standard pipelines very similar to those used for image classification.
To handle the varying duration of audio files, spectrograms were divided into 5-second chunks (with 50\% overlap) and treated as separate samples during training; classification decisions for a file are then averages over the chunks.

We tested the performance of algorithms from the two current leading families of deep learning algorithm: convolutional neural networks (CNNs) and transformers \citep{Stowell:2022review}.
For the transformer, we used an algorithm named PaSST that \cite{Ghani:2024} reported as strongly-performing for birdsong.
For the CNN, we used EfficientNetv2 \citep{Tan:2021}.

The EfficientNet algorithms are popular and robust CNNs.
Since our particular use of EfficientNet has not been documented in the academic literature, we give here a brief summary of the implementation, which we will refer to as `InsectEffNet'.

We selected EfficientNetv2 based on the outcome of a previous data contest which used the InsectSet66 dataset.\footnote{\url{https://www.capgemini.com/in-en/news/inside-stories/catching-the-ai-bug/}}
We used the EfficientNetV2-S model, which has around 20 million parameters.
We use version of the model publicly available in the PyTorch `timm' library which had been pre-trained using ImageNet21k. This image-based pre-training is a common convenience which accelerates the convergence of training for the network.
As input to the network we used mel spectrograms with 128 frequency bands (from 0.4 to 22 kHz), calculated from audio standardized to 44.1 kHz.

InsectEffNet and its training procedure were implemented using the PyTorch Lightning framework.
The source code for InsectEffNet is available online under the open-source MIT license.\footnote{\url{https://github.com/danstowell/insect_classifier_GDSC23_insecteffnet}}

As a comparison, we also trained and evaluated the Transformer-based PaSST method. The method was used identically as described in \cite{Ghani:2024} which uses an upper sampling rate of 32 kHz for audio. We adapted the audio chunk size to the same (5 seconds) as InsectEffNet, and used the same number of 128 mel bands.

To counter class imbalance, for both classifiers we applied custom class weights during training. For each class (species), the weighting was chosen as
\begin{equation*}
w_i = 1 - \frac{n_i}{\sum_{j=1}^K{n_j}}
\end{equation*}
where $n_.$ is the number of files in a given class and $K$ the number of classes. This down-weights the more common classes in the dataset, such that the training of machine learning should lead to more even per-class performance.

Data augmentation, which synthetically adds variety to the training data, is common in deep learning to improve generalization. The PaSST method uses additive noise and MixUp to the waveform data. Each augmentation had a 60\% chance of being applied to a given waveform. For InsectEffNet we used a similar procedure as in \cite{Faiss:2023}, applying additive noise and/or impulse responses (environmental reverberation and sound absorption) to the waveform data. Each augmentation had a 15\% chance of being applied to a given waveform.

As is common in deep learning for audio, both classifiers perform inference on a fixed short duration of audio, and so for a longer audio file multiple predictions must be made and then pooled across time. The default for InsectEffNet is to output the mean prediction per species (`mean-pooling'), while \cite{Ghani:2024} outputs the maximum prediction (`max-pooling'). We evaluated both options, for each classifier.

We did not perform hyperparameter optimization (tuning) for either of these algorithms, because the focus of our current work is to provide baseline standard results; instead, we make use of the tuning that each algorithm had received prior to our main test. 

To evaluate performance for our classifiers, we focus primarily on the `F1 score' evaluated on the InsectSet459 test set. The F1 score is the harmonic mean of the `precision' and `recall' measures. It is recommended as a more reliable measure than `accuracy' for a dataset which is highly unbalanced in the number of examples per category, which is the usual case for wildlife observation data. We measured the F1 score in aggregate, and also for each of the 459 species separately, which enables us to inspect how performance varies from the commonly-represented to the less-common species.

\section{Results}

\subsection{Dataset characteristics}

\begin{figure}[!p]
\includegraphics[width=0.95\linewidth]{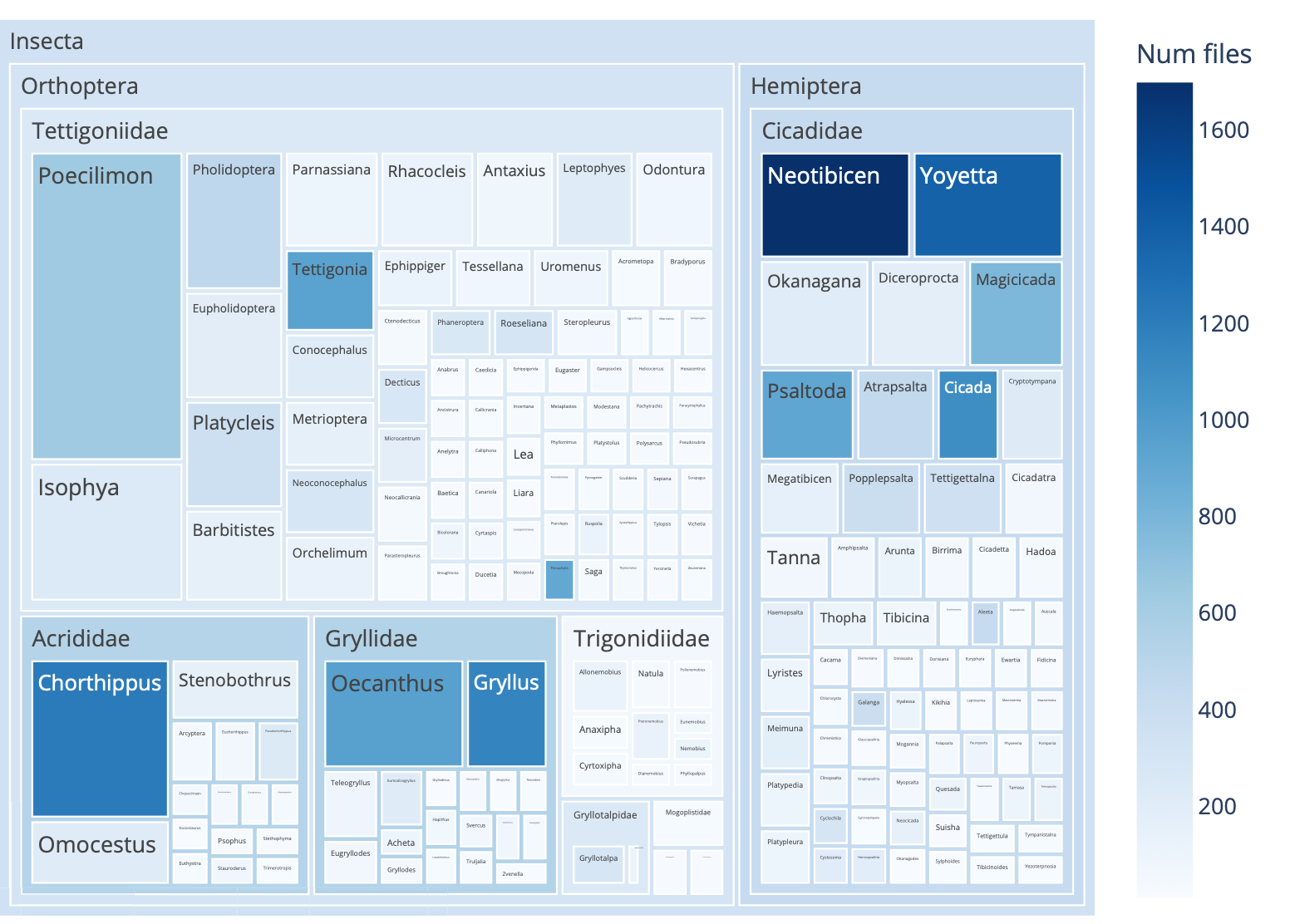}
\caption{Summary of the taxonomic groups represented within InsectSet459. The hierarchically-nested boxes represent taxonomic groups (some intermediate ranks such as subfamilies have been omitted for simplicity). For each box, the size indicates the number of species represented in the dataset, and the darkness of the box indicates the total number of sound recordings included.}
\label{fig:speciestreemap}
\end{figure}

The collected dataset covers 459 insect species: 310 Orthopteran species and 149 Cicada species (\FIG{speciestreemap}).
It contains 26399 files, with approximately equal amount of WAV and MP3 files (13120 WAV, 13279 MP3), and a wide range of sample rates, with more than 5000 sampled at rates higher than 50 kHz (\TABLE{filedurations}, \TABLE{fileformats}).

\begin{figure}[tp]
\includegraphics[width=1\linewidth]{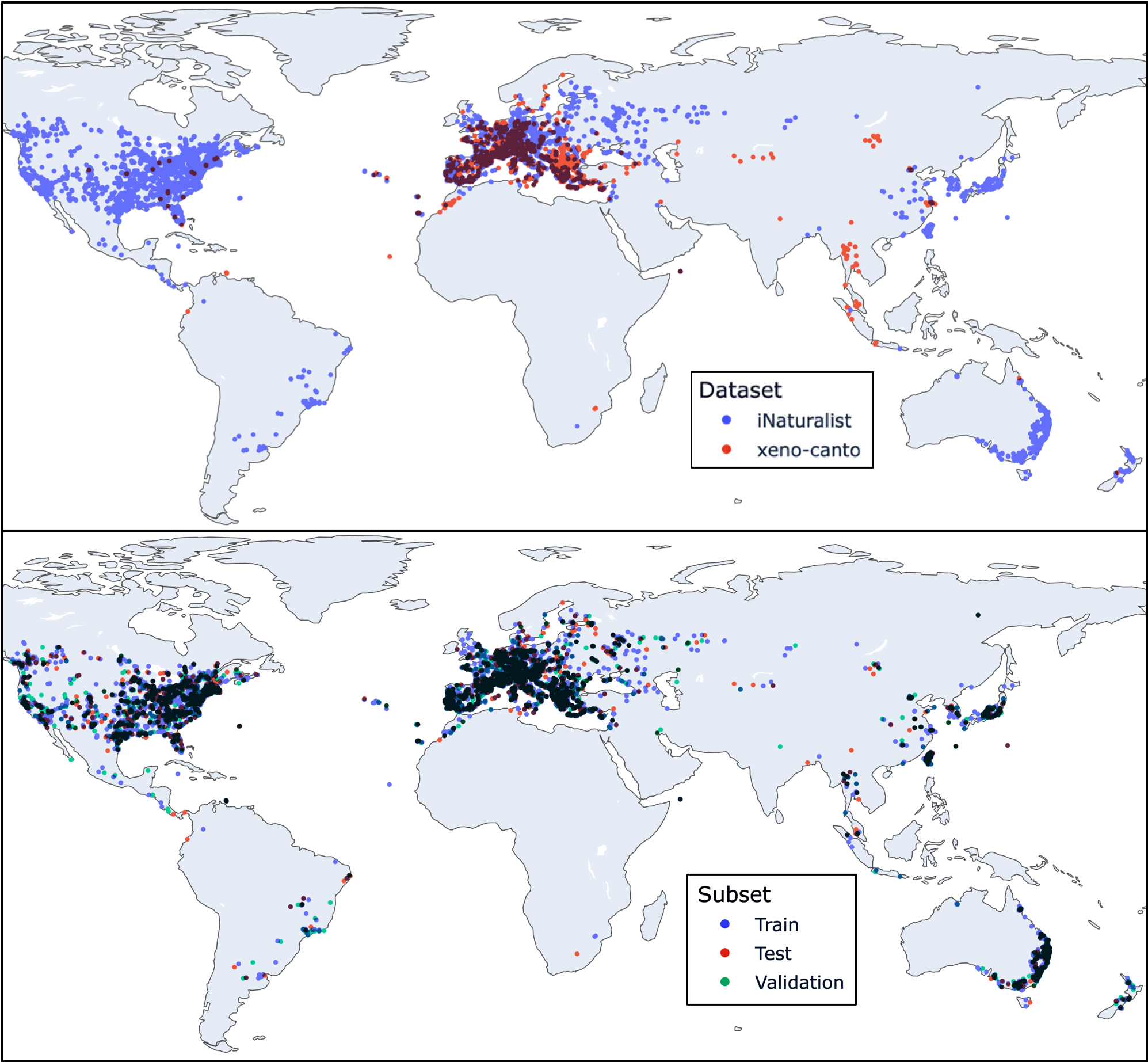}
\caption{The geographic locations of the files contained in InsectSet459. Out of 26399 files in the dataset, 305 do not have coordinates associated with them. (a) Upper plot: Recording locations split by source dataset. Blue dots indicate recordings sourced from iNaturalist, red dots show recordings from xeno-canto. Purple shows the overlap of both datasets.
(b) Lower plot: Recording locations split into the training, validation and test sets. Blue dots indicate recordings in the training set, red dots show recordings in the testing subset and green dots show recordings in the validation subsets. Other colors show the overlap between the subsets.}
\label{fig:datasets_subsets_map}
\end{figure}

\begin{figure}[pt]
\includegraphics[width=\linewidth]{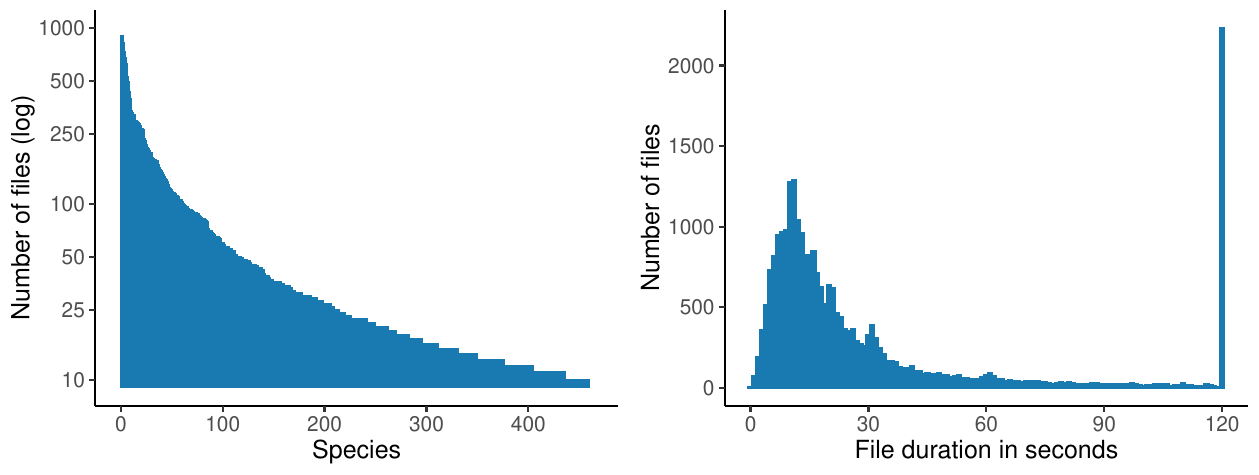}
\caption{a) Left plot: The number of files for all 459 species in the dataset, sorted in descending rank order and scaled logarithmically. This illustrates the strongly imbalanced nature of the data, a common aspect of bioacoustic datasets.
b) Right plot: The distribution of file durations in the dataset. Most files are around 10 seconds long. The large peak at 120 seconds is the result of trimming longer files to a maximum of two minutes.}
\label{fig:nfiles_durations}
\end{figure}

Of the 459 species, many are in the ``long tail'' of the data distribution with very few recordings per species: the most prevalent have more than 500 recordings, but around half have fewer than 25 (\FIG{nfiles_durations}). This level of class imbalance is a common aspect of datasets in ecology
\citep{Borowiec:2022}.

The geographic coverage of this dataset is heavily biased towards Europe and Northern America (\FIG{datasets_subsets_map}), likely due to the majority of contributors being located in these areas. It is possible that most, if not all known sound-producing Orthoptera and Cicadidae species from these areas are covered in this dataset. Tropical regions are heavily underrepresented, especially considering their much higher biodiversity among insects. This could also be explained by the large number of species in this area that are not known to produce sounds, or whose calls have not been described \citep{Riede:2024}.

\begin{table}[tp]
    \centering
    \begin{tabular}{lrr}
        Fold & Num files & Total duration (hours) \\
        \hline
        Train & 15873 & 137.3 \\
        Validation & 5307 & 46.2 \\
        Test & 5219 & 43.7 \\
        \hline
        Total & 26399 & 227.2
    \end{tabular}
    \caption{Sizes of the data folds in InsectSet459.}
    \label{tab:filedurations}

\end{table}

\begin{table}[tp]
    \centering


    \begin{tabular}{rr||rr||rr}
    Rate (kHz) & Num files & Rate (kHz) & Num files & Rate (kHz) & Num files\\
    \hline
8 & 91 & 48 & 4605 & 192 & 861 \\
11 & 3 & 49 & 2 & 200 & 38 \\
16 & 87 & 50 & 3 & 250 & 193 \\
22 & 58 & 63 & 11 & 256 & 639 \\
24 & 102 & 64 & 111 & 300 & 3 \\
26 & 3 & 88 & 28 & 313 & 297 \\
32 & 932 & 96 & 2145 & 320 & 2 \\
33 & 1 & 100 & 63 & 384 & 402 \\
38 & 11 & 125 & 269 & 500 & 4 \\
44 & 15432 & 152 & 3 \\
\\

    \end{tabular}

    \caption{Sample rates in InsectSet459, rounded to the nearest kHz.}
    \label{tab:fileformats}
\end{table}

\clearpage
\subsection{Performance of classifiers}


First, the hyperparameters of InsectEffNet were trained and tuned using the previous, smaller dataset InsectSet66. That dataset also offers a starting point to inspect the classification performance of InsectEffNet, before considering our main results from the larger dataset.
The performance was very strong, with an overall F1 score of 95.6\% and accuracy of 97.7\%. Although performance was best for the most commonly-represented species, it remained very high for most of the 66 species in this dataset (Appendix, \FIG{persp_classifiers_66}).
InsectEffNet achieved stronger performance than the previous work trained on IS66; that work was not aimed at maximizing performance, but investigating feature representations using a simple CNN architecture \citep{Faiss:2023}.
These initial results indicate that EfficientNet is a very well-performing CNN architecture, on this task as it has been for many other tasks reported in the literature.


We then performed a classification study using the new enlarged dataset IS459. We retrained InsectEffNet on our full-size dataset for 459 species.
We also trained and evaluated the transformer-based algorithm (named `PaSST') used in \cite{Ghani:2024}.
The deep learning classifiers attained F1 scores between 56\% and 58\%, with PaSST achieving the strongest results (\TABLE{classifiers_scores}). This level of performance is lower than achievable on IS66---as is to be expected for this substantially expanded task over 459 different species, including many species in the ``long tail'' of the data distribution with very few recordings per species (\FIG{nfiles_durations}).

The choice of temporal pooling did not modify InsectEffNet outcomes, possibly indicating its predictions were relatively consistent over the duration of an audio file. For PaSST, max-pooling gave a slightly better F1 score (\TABLE{classifiers_scores}).

\begin{figure}[tp]
\includegraphics[width=0.95\linewidth]{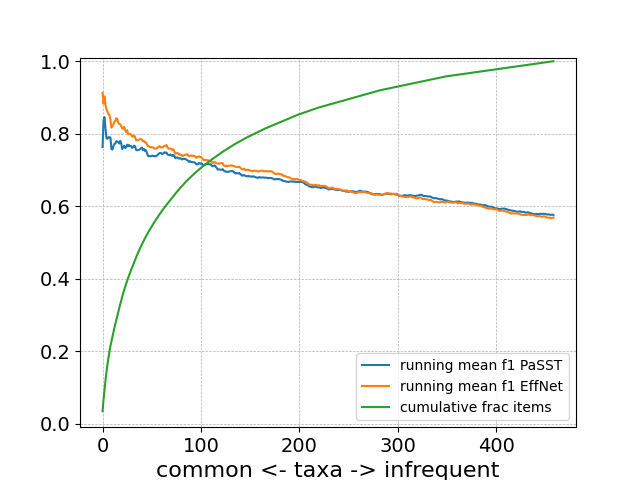}
\caption{Per-species classification performance of classifiers trained on IS459, evaluated on the test set. The species are ordered on the x-axis from most common to least common in the dataset, and for each one we plot a running mean of the F1 score---meaning the mean of the F1 score for all species that are equally or more common.}
\label{fig:persp_classifiers}
\end{figure}

The long tail effect can be seen directly when the IS459 classification results are plotted per species, in decreasing order of species frequency (\FIG{persp_classifiers}). The downward slope of the F1 curves indicates that the most common categories in IS459 can be classified by either algorithm with a performance above 80\%, but this rapidly decreases as the less-frequent categories are included.
For the infrequent categories, the per-species score has a very high variability (seen in Appendix \FIG{persp_classifiers_valtest}), which is to be expected when there are few examples available. Nevertheless, these infrequent categories are included so that the classifier can be exposed to a diverse variety of insect sounds. In some species, low performance could be caused by their frequency ranges lying outside the range of the spectrogram feature representations used in this work (InsectEffNet: up to 22 kHz, PaSST: up to 16 kHz), leaving little to potentially no species-specific information for the models to detect.

\begin{table}[tp]
    \centering
    \begin{tabular}{ll|rr}
    Model & Temporal pooling & F1 score (\%) & Accuracy (\%)\\
    \hline
    InsectEffNet     & Mean & 56.8 & 72.2 \\
    InsectEffNet     & Max & 56.8 & 72.2 \\
    PaSST     & Mean & 56.7 & 69.2 \\
    PaSST     & Max & 57.5 & 68.1 \\
    \end{tabular}
    \caption{Overall classification performance of classifiers trained on IS459, evaluated on the test set. F1 score is macro-averaged across classes; accuracy is averaged per-item.}
    \label{tab:classifiers_scores}
\end{table}

\section{Discussion}

In their systematic review, \cite{Kohlberg:2024} identified machine learning studies covering a total of 302 species for insect sound recognition. The InsectSet459 dataset thus represents an opportunity to substantially increase the coverage of bioacoustic machine learning in this domain by adding 406 species that have not been covered in machine learning studies according to the findings in \cite{Kohlberg:2024}. 
Our method was not designed to ensure complete species coverage for any given region; thus, using this dataset alone to train an automatic acoustic monitoring pipeline for a specific geographic region is not our recommended strategy, since all species from a given region are not necessarily represented. Instead, InsectSet459 could be useful for developing and testing machine learning methods specifically optimized for insect sounds and datasets with highly variable sample-rates. Its total size and number of species could realistically simulate a comprehensive dataset that covers all sound-producing species of a large geographic region (especially in non-tropical areas). Therefore, methodologies that are developed and work well on this dataset are likely to function well when used to acoustically monitor soniferous insect species in a real environment. A further step towards that goal would be to use InsectSet459 to pre-train models and then fine-tune with a smaller, perhaps strongly labeled dataset that covers a geographic region, taxonomic group or a specific task on which a model should be deployed.

The annotation file for InsectSet459 contains weak labels, meaning that each audio sample is only associated with one species-label, without on- and offset times or durations annotated. Therefore, this dataset by itself is not suitable for training sound-event detection algorithms that predict precise on- and offsets in longer recordings. But pre-training with InsectSet459 and fine-tuning with a small, strongly-labeled dataset could be beneficial and reduce the amount of precise labeling work that would need to be done otherwise. Some recordings are annotated with additional labels that list species that are audible in the background. This could be useful for certain applications, but background labels are only available for a subset of the data and many recordings contain species in the background which are not labeled.


Our curation of InsectSet459 benefits from the growth in insect-focused contributions to community bioacoustics collections (xeno-canto and iNaturalist), which were also the source for our prior InsectSet66. InsectSet459 is much larger in the number of insect species represented, thus having much better potential for complete species coverage in selected locations.
IS459 is more than 9 times larger than IS66 in total audio duration (IS66: 24 hours; IS459: 227 hours). This makes the dataset more useful for data-hungry applications such as deep learning. Regarding total data volume, note that we limited individual audio file durations to a maximum of 2 minutes. Without this limit, the dataset could be much larger still, but with a kind of false or unhelpful replication, with limited benefits. We chose instead to focus on increasing dataset size through diversity of audio samples. 

We did not manually edit the audio files to trim them down to song bouts (echemes), as was done for IS66: instead we trimmed them to automatically-selected 2-minute segments. This is done to create a curation process that scales to large data sizes. It risks making some imperfect edits, which could be unhelpful for algorithm training. On the other hand it also corresponds more closely to automated passive acoustic monitoring, in which arbitrary sound segments are usually submitted for analysis.

An important goal in this work was also to preserve ultrasonic information where available (in around 25\% of our data), since it can be an important acoustic component for various species and is a potential additional information source for recognition. This makes a distinctive characteristic of IS459 intended to improve automatic insect recognition in particular.

\subsection{InsectSet459 compared against iNatSounds}


An interesting point of comparison is against the recently-released \textit{iNatSounds} dataset \citep{Chasmai:2024}.
That dataset contains a similar quantity of files, from a common data source, but with a much wider taxonomic span: over 5,500 species, including insects. However, IS459 covers many more insect species: iNatSounds has only 111 Insecta in its test set.
Further, the pre-processing for iNatSounds down-sampled data to a common 22 kHz. This standardization helps to simplify the workflow for users, but reduces the options for enabling high-fidelity analysis and machine learning for insects in particular, given the very wide variation of frequency bands used by insects.
We used both Xeno-canto and iNaturalist data to add further diversity of source material. Compared to iNaturalist alone, this did not expand the geographic range as much as we had expected, since both collections contain a large fraction of material from common locations in Europe (\FIG{datasets_subsets_map}). However, the union of the two data sources served to increase coverage of habitats in north Africa and central and southern Asia.


A further difference is that we chose to ensure fully-open licensing, to maximize the re-usability of the dataset. This led to some different decisions of curation: we excluded audio files with no-derivatives or non-commercial requirements.
This dataset can be freely shared and republished by third parties, under the terms of the Creative Commons 4.0 Attribution license.
By contrast, iNatSounds is made available under a ``terms of service'' agreement with conditions of no commercial use, and no permission to redistribute, presumably because of considerations about the licensing of some of the included content.



The above considerations mean that there are some situations in which IS459 will be especially useful, and some in which iNatSounds is preferred.
We aim for IS459 to maximally useful to increase the quality of insect sound recognition, and to be freely distributable. Yet there may also be many situations in which users wish to include insect recognition into multi-taxon soundscape classifiers.

An important consideration: most users \textbf{should not combine IS459 with iNatSounds}, because there is overlap between them. In particular, users should not use iNatSounds training when evaluating on IS459, because there is no clear separation between iNatSounds and our testing set. It is likely that iNatSounds contains audio data that can also be found in our testing set, and thus that iNatSounds may lead to falsely-inflated estimates of performance. Users who wish to make joint use of the datasets can refer to IS459 annotation file which contains the observation link as well as the direct link to each of our source files. Alternatively, the insect records could be excluded from iNatSounds before combining. 

\subsection{Automatic classification results}


In our baseline classification experiment, we found that two very different deep learning classifiers attained a similar level of performance: around 57\% F1-score overall, rising to more than 80\% for the most common classes.
As an indicative comparison: for a similar task with focal birdsong recordings, the BirdNet classifier was reported to show an overall accuracy score of 77.7\% over 984 species, whereas in our case accuracy reaches 72.2\% \citep{Kahl:2021}.
Performance may thus already be good enough for some deployed use, especially for common species, although this should be undertaken with careful piloting.


There is a general decline in performance for the species for which we have a low number of examples. This decline was expected, given the dataset characteristics, since it is a common phenomenon in machine learning. For some data-poor categories, the F1 score per class was observed very low indeed, below 20\% for both classifiers (Appendix, \FIG{persp_classifiers_valtest}).
For ecosystem monitoring, the most important future improvements on these baseline results should be those that improve recognition for the infrequent cases. This can come from innovations in machine learning, but of course would also benefit from additional data samples in community open data collections.

For the information given to the baseline classifiers, we limited attention to spectrograms in the audible frequency range (up to 22 kHz). This likely affected performance, and could be an explanation for the poor performance on certain species. Future work on multi-sample-rate models could thus improve performance especially for ultrasonic species.

Regarding the machine learning algorithms, it is intriguing that the overall performance scores of the two models converge to very similar values, given that the underlying algorithms are very different from one another.
In fact, other implementation decisions have an equal magnitude of impact, such as the decision whether to mean-pool or max-pool an algorithm's predictions over time.
It is unlikely that these different models extract the same features, yet the convergence implies they might both be extracting a very large fraction of the information that they can reasonably get from the audible-range mel-spectra we use. This convergence is not complete---as can be seen in the variation in per-class scores---leaving room for improvements even without modifying the input representation.
But, again, we suggest that improving the input representation will also be of interest, and have designed IS459 to enable that.

\subsection{Next steps in machine learning for insect sound surveying}

It is important that further work is conducted to develop reliable automatic recognition for insect sounds. We hope that IS459 will facilitate this. In particular we expect and recommend that future work should make use of the ultrasound component of many insect sounds, which could include the possibility for multi-rate machine learning which is not restricted to a single sample rate.
Such methods might sidestep the spectrogram computation pre-processing, which in our baseline systems we have not questioned.
Additionally, in recent years self-supervised learning has been demonstrated to create very good automatic recognition systems, including for animal sounds \citep{Schafer:2024animal2vec, hagiwara:2023}. We have not investigated that in the present paper, but we expect self-supervised learning to be useful in uncovering general representations for insect sound.

Our dataset shares with all collections the limitation of incomplete coverage of insects and their acoustic behavior. Insect sounds may be simple compared against those of songbirds and other animals; however, they vary due to atmospheric conditions (especially temperature), and are often heard in `choruses', neither of which are addressed by the remit of this dataset. We strongly encourage the collection of more diverse open data as evidence of insect acoustic behavior. The work we have presented may even help with this, as a step on the way to developing automatic systems that can distinguish the similar and dissimilar, the known and unknown, in insect sound.

Taken together, there are strong prospects that automatic recognition of insect sounds in the wild can reach a level of reliability high enough to be used in large-scale monitoring projects. Open datasets and open development of methods will continue to be important in this undertaking.

\section{Acknowledgments}

We thank the many contributors to the xeno-canto and iNaturalist collections, for sharing their recordings, without whom this work would not have been possible.
We highlight the following users who each contributed over 500 recordings: Baudewijn Od\'{e} (also for project advice), Christie (iNaturalist username), Joel Poyitt, K.-G. Heller, S. Ingrisch, Cedric Mroczko.


The InsectEffNet classifier was developed as part of a CapGemini ``Global Data Science Challenge'', supported by Amazon Web Services.
The classifier code was contributed by a CapGemini team and authored by Raffaela Heily, Lukas Kemetinger, Dominik Lemm and Lucas Unterberger. It is used here with permission.

MF was supported by the EU MSCA Doctoral Network Bioacoustic AI (BioacAI, 101071532). BG was supported by EU-funded Horizon projects MAMBO (101060639), GUARDEN (101060693) and TETTRIs (101081903).

\bibliography{is459refs}

\clearpage
\newpage
\appendix
\section{Classifier performance: additional plots}

\subsection*{InsectSet66:}

\begin{figure}[hp]
\includegraphics[width=0.75\linewidth,page=2]{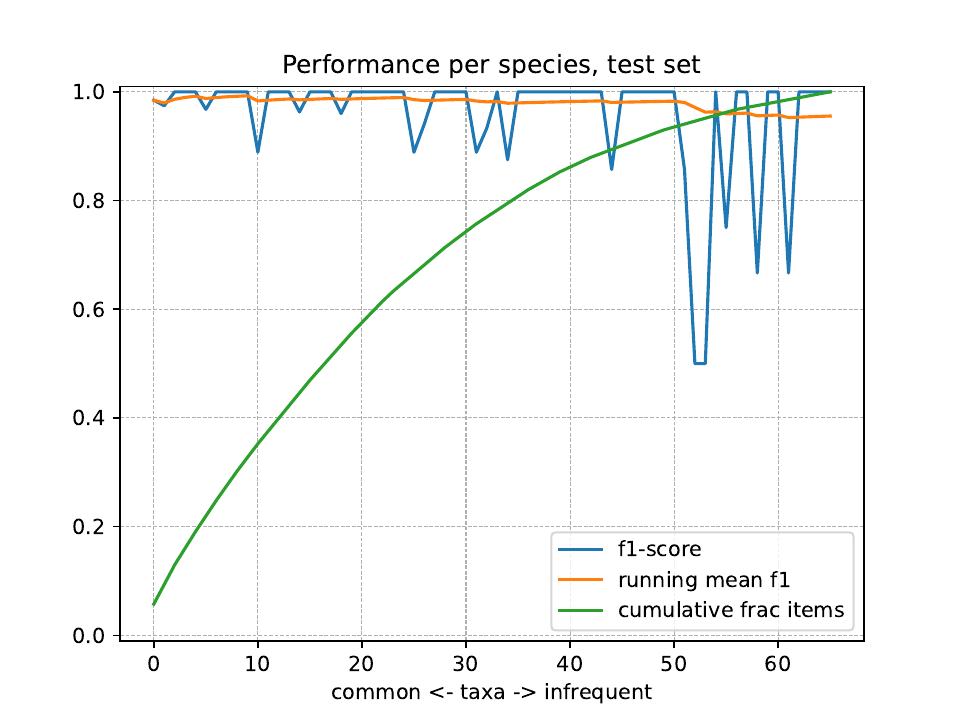}
\caption{Per-species classification performance of InsectEffNet trained on InsectSet66, evaluated on the test set. The species are ordered on the x-axis from most common to least common in the dataset, and for each one we plot a running mean of the F1 score---meaning the mean of the F1 score for all species that are equally or more common.}
\label{fig:persp_classifiers_66}
\end{figure}

\clearpage
\subsection*{InsectSet459:}

\begin{figure}[hp]
\includegraphics[width=0.45\linewidth,page=1]{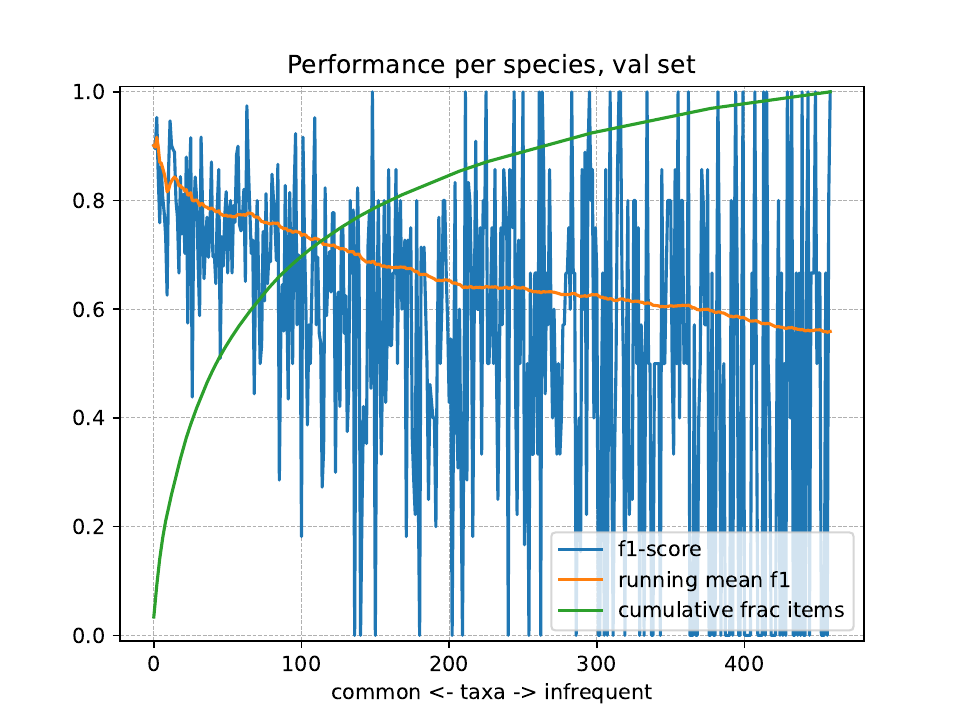}
\includegraphics[width=0.45\linewidth,page=2]{plot_performance_ranked_persp.pdf}

\includegraphics[width=0.45\linewidth]{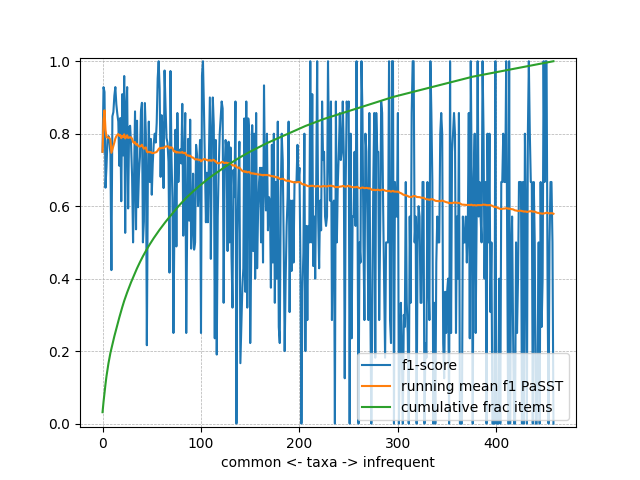}
\includegraphics[width=0.45\linewidth]{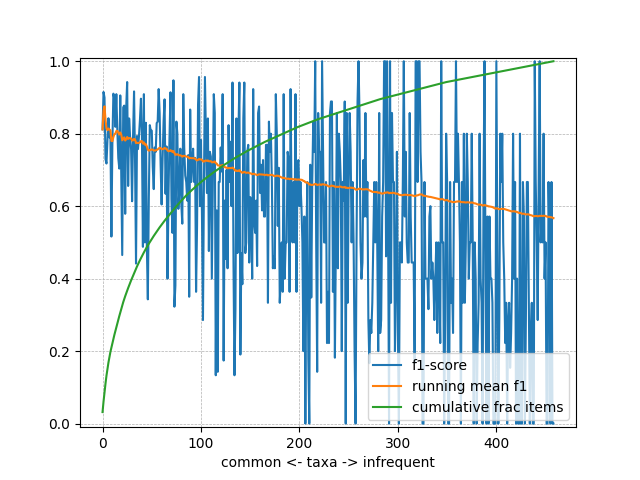}
\caption{Per-species classification performance of InsectEffNet and PaSST trained on InsectSet459, evaluated on the validation and test sets. The species are ordered on the x-axis from most common to least common in the dataset, and for each one we plot the F1-score and a running mean. a) Top left: Performance of InsectEffNet on the validation set. b) Top right: Performance of InsectEffNet on the test set. c) Bottom left: Performance of PaSST on the validation set. b) Bottom right: Performance of PaSST on the test set.
}
\label{fig:persp_classifiers_valtest}
\end{figure}

\end{document}